\documentclass[aps,pre,twocolumn,groupedaddress,showpacs]{revtex4}
\usepackage{graphicx}

\def\x{{\bf x}}
\def\y{{\bf y}}

\def\R{{\bf R}}
\def\0{{\bf 0}}

\begin{document}

\title{Moments of vicious walkers and M\"obius graph expansions}

\author{Makoto Katori}
\email[]{katori@phys.chuo-u.ac.jp}
\author{Naoaki Komatsuda}
\email[]{komatuda@phys.chuo-u.ac.jp}
\affiliation{
Department of Physics,
Faculty of Science and Engineering,
Chuo University, Kasuga, Bunkyo-ku, Tokyo 112-8551, Japan}

\date{March 31, 2003}

\begin{abstract}
A system of Brownian motions in one-dimension all started from 
the origin and conditioned never to collide with each other 
in a given finite time-interval $(0, T]$ is studied.
The spatial distribution of such vicious walkers can be
described by using the repulsive eigenvalue-statistics of
random Hermitian matrices and it was shown that the present
vicious walker model exhibits a transition
from the Gaussian unitary ensemble (GUE) statistics to
the Gaussian orthogonal ensemble (GOE) statistics as the time
$t$ is going on from 0 to $T$.
In the present paper, we characterize this GUE-to-GOE transition 
by presenting the graphical expansion formula for the
moments of positions of vicious walkers. 
In the GUE limit $t \to 0$, only the ribbon graphs 
contribute and the problem is reduced to the classification
of orientable surfaces by genus. Following the time evolution
of the vicious walkers, however, the graphs with twisted ribbons,
called M\"obius graphs, increase their contribution
to our expansion formula, and we have to deal with the
topology of non-orientable surfaces. Application of the
recent exact result of dynamical correlation functions
yields closed expressions for the coefficients in the
M\"obius expansion using the Stirling numbers of the
first kind.

\end{abstract}


\maketitle

\section{Introduction} 

Statistics of a set of one-dimensional random walks 
{\it conditioned never to collide} in a given time interval, say $T$,
has its own importance in statistical physics,
since, if we set $T \to \infty$, it realizes the one-dimensional
Fermi statistics \cite{deG68}, and with $T < \infty$ it is used
to analyze the models for wetting and melting phenomena
\cite{Fis84}. We will refer to such non-colliding random walks
and the continuum counterpart, non-colliding Brownian motions,
simply as {\it vicious walks} following the terminology 
used by M. Fisher \cite{Fis84}. For the pioneering work
on vicious walker models, see
\cite{HF84,Fold,AME91,MB93,EG95}.
A generic setting of vicious walk problems is discussed in
\cite{CK03}.
Recently the interest on the vicious walks 
in mathematical physics is renewed and growing very rapidly,
for close relationships of the vicious walk problem 
with the study of ensembles of Young tableaux 
and the symmetric functions \cite{GOV98,KGV00,KGV02}, 
the theories of orthogonal polynomials and random matrices 
\cite{Bai00,NF02}, and some topics of representation theory and 
probability theory \cite{Gra99,OY02,KT02a} have been clarified.

In an earlier paper \cite{KT02b}, the continuum limit of non-colliding 
random walks on a lattice was taken by letting
the temporal and spatial units $\Delta t, \Delta x$ go to zero
with the relation $\Delta t \propto (\Delta x)^2$
and a system of non-colliding Brownian motions was derived.
Since each random walk tends to be a Brownian motion
in this {\it diffusion scaling limit}, such a construction
of non-colliding Brownian motion is plausible, and indeed
mathematical rigor can be established as 
a functional central limit theorem of vicious walks \cite{KT02a}.
The important fact is that the repulsive interaction among
the obtained Brownian particles is no longer contact
interaction as in the original vicious random walks on a lattice
but is long ranged. The origin of this long-ranged
interaction is the restriction of allowed configurations by the
non-colliding condition, so that we can say it is an
example of entropy-origin effective force.
Moreover, the following setting was considered in our vicious
walks; (i) even after taking the continuum limit,
still we let the time interval, in which the non-colliding
condition is imposed, be finite $T < \infty$, and
(ii) all the Brownian particles are assumed to start
from an origin and the time interval of non-colliding
is $(0, T]$.

In this setting the process is temporally inhomogeneous.
At the very early stage $t \ll T$ the repulsive interaction
should be strong, since the Brownian motions will be
restricted so that they will not collide for a long
time period up to time $T$ in the future.
As the time $t$ going on, the strength of repulsion is
decreasing as is the remaining time until $T$,
and attains its minimum at the final time $t=T$,
at which there is no more restriction of motion
in the future $t > T$. It was Dyson's idea that such systems
of Brownian motions with long-ranged repulsion could have
the equilibrium states, which can be described by the
distribution functions of eigenvalues of
random matrices in the ensembles appropriately specified
by symmetry of the system \cite{Dys62}.
The previous paper \cite{KT02b} showed that the spatial 
distribution of vicious walkers at $t \ll T$ realizes
the eigenvalue-statistics of the Gaussian unitary ensemble 
(GUE), one at $t=T$ does that of the Gaussian orthogonal 
ensemble (GOE), and the GUE to GOE transition is 
observed in the intermediate time, 
which is equivalent with the transition 
studied in the two-matrix model by 
Pandey and Mehta \cite{PM83,Meh91}.

In the present paper, we will characterize this transition
in distribution by calculating the moments of the
positions of $N$ particles as functions of time $t$,
\begin{equation}
M_{N,T}(t,k)= 
\left\langle \sum_{j=1}^{N} x_{j}^{2k} 
\right\rangle_{t}, \quad
k=1,2,3, \cdots, 
\label{eqn:moment0}
\end{equation}
where $x_{j}$ denotes the position of $j$-th vicious walker
and $\langle \ \cdot \ \rangle_{t}$
the average at time $t \in (0, T]$.
In addition to such an interest of statistical physics,
we can put emphasis that our present study possesses another
importance as an interesting application of the graphical 
expansion theory.
In high energy physics, graphical expansions for the matrix
models of SU($N$) gauge theory were studied
using {\it ribbon graphs} representing propagators
and it was shown that the dominant graphs 
for large $N$ are the planar ones and 
the leading corrections come from the graphs
embedded in a torus, which are depressed by a factor
$1/N^2$ with respect to the planar graphs \cite{tHo74,BIPZ78}.
On the other hand, it was clarified that in the graphical
expansion of SO($N$) gauge theory the leading corrections 
only are depressed by a factor $1/N$ for large $N$
with respect to the dominant planar graphs, in which propagators
can be represented by {\it twisted ribbons} \cite{Cic82,BN91}.
In the gauge theory, the presence of non-Gaussian
interaction terms of cubic or higher power is crucial and 
it leads to {\it triangulations of random surfaces}.
Although the Gaussian matrix models associated with 
the present vicious walker model are not related with
the triangulation problem of random surfaces and
only give purely enumerative problems of surfaces, 
the proper structures of large $N$ expansions
in SU($N$) and SO($N$) gauge theories are also 
found in the GUE and GOE models, respectively.
We can show that, if $T \to \infty$ in our vicious walker model,
the moments of the walker positions (\ref{eqn:moment0})
can be calculated by the graphical expansion method
of GUE using the ribbon graphs and the results are
given in the form of power series in the inverse of 
the {\it square} of matrix size $N$,
$$
M_{\rm GUE}(k) \propto N^{k+1} \sum_{g \geq 0} \varepsilon_{g}(k)
\left(\frac{1}{N^2} \right)^{g}.
$$
Here the coefficients $\varepsilon_{g}(k)$ are the
numbers of {\it orientable surfaces of genus} $g$ made from
$2k$-gon by some specified procedure
\cite{HZ86} (see also \cite{Zvo97} and Sec.5.5 in
\cite{Meh91}).
On the other hand, for the GOE
we have to take into account the {\it non-orientable
surfaces} as well as the orientable ones and
the expansion is in the form of power series in $1/N$. 
In other words, in order to generate necessary surfaces,
we need to use twisted ribbons, whose type of
graphs is now called of {\it M\"obius graphs} \cite{Sil97,MW02}.
For the moments (\ref{eqn:moment0}) of our vicious walkers,
we will perform the M\"obius graph expansion
in the present paper. Now
the weights of the graphs are depending on the time
$t$; in the limit $t \to 0$ all the weights on the 
twisted ribbons are zero, but they are growing as time
$t$ going on, and at $t=T$ twisted ribbons are equally
weighted as untwisted ones.
This gives another characterization of the temporally
inhomogeneity of the process and the GUE-to-GOE
transition.

Quite recently Nagao, Tanemura and one of the present authors
applied the method of skew orthogonal polynomials
and quaternion determinants developed for the multimatrix
models \cite{NF99,FNH99,Nag01}
to the vicious walk problem and derived the quaternion determinantal
expressions for dynamical correlation functions \cite{NKT03}.
Using this result, we will present an expression
of the coefficients in our expansion of moments
using the Stirling number of the first kind.

The paper is organized as follows. In Sec.II we briefly
review the previous results reported in \cite{KT02b}
and give the precise definition of the moments which
we will study. Graphical representations are demonstrated
in Sec.III. Application of the result of \cite{NKT03}
is given in Sec.IV to give the expression
for the coefficients of expansion and some concluding remarks
are given in Sec.V.
Appendices A and B are prepared to derive the expression of the
density function used in Sec.IV and the $1/N$-expansions
of the one-point Green function discussed in Sec.V, respectively.

\section{VICIOUS WALKS AND MATRIX MODEL}

We study a continuum model of vicious walks,
the non-colliding Brownian motions in the time interval $(0,T]$,
constructed in \cite{KT02a,KT02b} as the diffusion scaling limit of 
vicious random walks on a lattice. First we briefly review our
previous results.
The configuration space of the present $N$ vicious walkers is
${\bf R}_{<}^{N}= \{\x=(x_{1}, x_{2}, \cdots, x_{N}) \in
\R^N; x_{1} < x_{2} < \cdots < x_{N}\}$,
where $\R$ is a set of all real numbers.
The probability density of vicious walkers at time 
$t \in (0, T]$ with the initial condition that all walkers
start from the origin $\0$ is denoted by $\rho_{N, T}(t, \x)$.
It was given as
\begin{equation}
\rho_{N, T}(t, \x)= C e^{-|\x|^2/2t} h_{N}(\x)
{\cal N}_{N}(T-t, \x)
\label{eqn:P0NT1}
\end{equation}
with 
$C= 2^{-N/2} T^{N(N-1)/4} t^{-N^2/2}
/\prod_{j=1}^{N} \Gamma(j/2)$
for $\x \in \R_{<}^{N}$, where
$\Gamma(z)$ is the Gamma function,
$h_{N}(\x)=\prod_{1 \leq j < \ell \leq N} 
(x_{\ell}-x_{j})$
and
$$
{\cal N}_{N}(s, \x)=\int_{\R^{N}_{<}} d \y \
\det_{1 \leq j, \ell \leq N} \left(
\frac{1}{\sqrt{2 \pi s}}
e^{-(x_{j}-y_{\ell})/2s} \right).
$$
By using the Harish-Chandra/Itzykson-Zuber integral formula
\cite{HC57,IZ80,Meh81}, we will see that
$\rho_{N, T}(t, \x)$ is proportional to
the integral
\begin{equation}
h_{N}(\x)^2 \int dU \int dA \exp \left(
-{\rm tr} {\cal H}(U^{\dagger} X U, A) \right)
\label{eqn:P0NT2}
\end{equation}
with
$$
{\cal H}(H, A)=\frac{T}{2t(T-t)} H^2 -
\frac{T}{t(T-t)} HA + \frac{T^2}{2t^2(T-t)} A^2,
$$
where $X$ is the $N \times N$ diagonal matrix with
$X_{j \ell}=x_{j} \delta_{j \ell}$, 
and the integrals $\int dU$ and
$\int dA$ are taken over the groups of $N \times N$ unitary
matrices $\{U\}$ and real symmetric matrices $\{A\}$, 
respectively.
The proportional coefficient is determined so that
the probability density is normalized.
On the other hand, the integral over $A$ in (\ref{eqn:P0NT2})
can be regarded as the convolution of the Gaussian distribution
of complex Hermitian matrices $H$ with variance $t(1-t/T)$
and that of real symmetric matrices $A$ with variance $t^2/T$,
and thus we have the expression
\begin{equation}
\rho_{N, T}(t, \x) \propto h_{N}(\x)^2 \int dU \
\mu_{N, T}(t, U^{\dagger}XU) 
\label{eqn:P0NT3}
\end{equation}
with
\begin{eqnarray}
&& \mu_{N, T}(t, H) \nonumber\\
&& \quad \propto
\exp \left( - \sum_{j, \ell} \left\{
\frac{(H_{j \ell}^{\rm R})^2}{2t}
+\frac{(H_{j \ell}^{\rm I})^2}{2t(1-t/T)} \right\} \right),
\label{eqn:mu}
\end{eqnarray}
where and in the following we use the abbreviations
$z^{\rm R}$ and $z^{\rm I}$ for the real and the imaginary
part of the complex number $z$, respectively,
i.e. $z=z^{\rm R}+i z^{\rm I}$
for $z \in {\bf C}$ with $i=\sqrt{-1}$.
Remark that, if we set
\begin{equation}
  c=\sqrt{\frac{t(2T-t)}{T}}
\label{eqn:c}
\end{equation}
and $\alpha^2=1-t/T$,
$c^{N} \mu_{N,T}(t, c H)$ is equal to the probability density
of the two-matrix model of Pandey and Mehta
with the parameter $\alpha$ \cite{PM83,Meh91}.
Corresponding to changing the parameter $\alpha$ from 1 to 0
in the Pandey-Mehta two-matrix model,
a GUE-to-GOE transition occurs in the time
development of particle distribution in our
vicious walks.

Now we define the quantity, which we will study in the present
paper; the moment of particle positions
in the vicious walks. Since the distribution 
(\ref{eqn:P0NT1}) of $\x$ is symmetric about the origin $\0$,
all of the odd moments vanish. The even moments are defined
and denoted as follows,
\begin{eqnarray}
M_{N,T}(t, k) &=& \left\langle \sum_{j=1}^{N} x_{j}^{2k}
\right\rangle_{t} \nonumber\\
&=& \int_{\R_{<}^{N}} d \x \
\sum_{j=1}^{N} x_{j}^{2k} \rho_{N, T}(t, \x)
\label{eqn:Mk1}
\end{eqnarray}
for $k=1,2, \cdots$.

\section{GRAPHICAL EXPANSIONS}

\subsection{Wick formula}

First we notice that (\ref{eqn:P0NT3}) is invariant under any
permutation of $x_{1}, x_{2}, \cdots, x_{N}$. Then (\ref{eqn:Mk1})
is written as
\begin{eqnarray}
&& M_{N, T}(t, k)
\propto \frac{1}{N !} \int_{\R^N} d \x \ h_{N}(\x)^2
\nonumber\\
&& \quad 
\times \int dU \sum_{j=1}^{N} x_{j}^{2k} \mu_{N, T}
(t, U^{\dagger} X U). 
\label{eqn:Mk2}
\end{eqnarray}
Next we introduce the integration measure for the 
$N \times N$ complex Hermitian matrices
$$
dH = \prod_{1 \leq j \leq \ell \leq N} 
d H_{j \ell}^{\rm R}
\prod_{1 \leq j < \ell \leq N} 
d H_{j \ell}^{\rm I}.
$$
Since $dH \propto dU \times h_{N}(\x)^2 d \x$,
if $\x=(x_{1}, \cdots, x_{N})$ are the eigenvalues of
$H$ and $d\x=\prod_{j=1}^{N} dx_{j}$ 
({\rm e.g.} see \cite{Meh91}), and
$\sum_{j=1}^{N} x_{j}^{2k}={\rm tr} H^{2k}$ for
$H=U^{\dagger} X U$ with any unitary matrix $U$, 
(\ref{eqn:Mk2}) with (\ref{eqn:mu}) becomes
\begin{equation}
M_{N,T}(t, k)= \left\langle {\rm tr} H^{2k} \right\rangle,
\label{eqn:Mk3}
\end{equation}
where 
$
    \langle \ f \ \rangle =
    \int dH \ f \ \mu_{N, T}(t,H) 
$
for functions $f$ of the elements of $H$
with
\begin{eqnarray}
\mu_{N, T}(t, H) &=& \prod_{j=1}^{N} 
\frac{e^{-(H_{jj}^{\rm R})^2/2t}}{\sqrt{2 \pi t}}
\prod_{1 \leq j < \ell \leq N}
\frac{e^{-(H_{j \ell}^{\rm R})^2/t}}{\sqrt{\pi t}}
\nonumber\\
&& \nonumber\\
&\times&
\prod_{1 \leq j < \ell \leq N}
\frac{e^{-(H_{j \ell}^{\rm I})^2/t(1-t/T)}}
{\sqrt{\pi t(1-t/T)}}.
\label{eqn:mu2}
\end{eqnarray}
Note that
$$
  {\rm tr}(H^{2k})=\sum_{j_{1}, j_{2}, \cdots, j_{2k}}
   H_{j_{1} j_{2}} H_{j_{2} j_{3}} \cdots
H_{j_{2k-1} j_{2k}} H_{j_{2k} j_{1}},
$$
where the sum is taken over all $N^{2k}$ combinations
of indices $j_{1}, j_{2}, \cdots , j_{2k}$, and that 
$H_{j \ell}=H_{j \ell}^{\rm R}+i H_{j \ell}^{\rm I}$ with
the Hermitian condition 
$H_{\ell j}^{\rm R}=H_{j \ell}^{\rm R},
H_{\ell j}^{\rm I}=-H_{j \ell}^{\rm I}$,
the integrand ${\rm tr}H^{2k}$ in (\ref{eqn:Mk3}) is
a polynomial of the $N^2$ independent random variables
$\{H_{j \ell}^{\rm R}; 1 \leq j \leq \ell \leq N \} \cup 
\{H_{j \ell}^{\rm I}; 1 \leq j < \ell \leq N \}$.
Since the probability density (\ref{eqn:mu2}) is a
product of independent Gaussian integration-kernels,
we can apply the Wick formula with the variances
\begin{eqnarray}
&&\langle (H^{\rm R}_{j \ell})^{2} \rangle 
= \frac{t}{2}(1+\delta_{j \ell}), \quad
\langle H^{\rm R}_{j \ell} H^{\rm I}_{j \ell} \rangle= 0,
\nonumber\\
&& 
\langle (H^{\rm I}_{j \ell})^{2} \rangle
= \frac{t(1-t/T)}{2} (1-\delta_{j \ell}),
\label{eqn:Wick1}
\end{eqnarray}
for $1 \leq j \leq \ell \leq N$, 
where $\delta_{j \ell}$ is 
Kronecker's delta.

We can readily prove that 
(\ref{eqn:Wick1}) is equivalent with 
\begin{equation}
  \langle H_{j \ell} H_{m n} \rangle
= \frac{c^2}{2} \left( \delta_{j n} \delta_{\ell m}
+ \gamma \, \delta_{j m} \delta_{\ell n} \right),
\label{eqn:Wick2}
\end{equation}
where $c$ is given by (\ref{eqn:c}) and
\begin{equation}
  \gamma= \frac{t}{2 T-t}.
\label{eqn:gamma}
\end{equation}
The Wick formula for (\ref{eqn:Mk3}) is thus
\begin{widetext}
\begin{eqnarray}
M_{N,T}(t, k) &=& \sum_{j_{1}, j_{2}, \cdots, j_{2k}}
\sum_{\pi \in S_{2k}:{\rm R}}
\langle H_{j_{\pi(1)} j_{\pi(1)+1}} H_{j_{\pi(2)} j_{\pi(2)+1}} 
\rangle
\langle H_{j_{\pi(3)} j_{\pi(3)+1}} H_{j_{\pi(4)} j_{\pi(4)+1}} 
\rangle \nonumber\\
&& \hskip 5cm
\cdots
\langle H_{j_{\pi(2k-1)} j_{\pi(2k-1)+1}} 
H_{j_{\pi(2k)} j_{\pi(2k)+1}}
\rangle,
\label{eqn:Mk4}
\end{eqnarray}
\end{widetext}
with the identification $j_{2k+1}=j_{1}$, 
where the first sum is taken over all $N^{2k}$ combinations
of indices $j_{1}, j_{2}, \cdots, j_{2k}$,
and the second one over the set of permutations $S_{2k}$ 
of $\{1,2, \cdots, 2k\}$ with the restriction
\begin{eqnarray}
{\rm R}: &&
\pi(1) < \pi(3) < \cdots < \pi(2k-1), \nonumber\\
&& \pi(2j-1) < \pi(2j), \quad 1 \leq j \leq k.
\nonumber
\end{eqnarray}
The total number of the terms in the second summation is
$(2k-1)!!$.

\subsection{An example: the fourth moment}

In this section, by performing calculation of the
fourth moment $M_{N,T}(t, 2)=\langle {\rm tr} H^4 \rangle$,
we will demonstrate how to obtain graphical expansions
from the Wick formula (\ref{eqn:Mk4})
with the variance (\ref{eqn:Wick2}).
We start from the Wick formula for $M_{N,T}(t, 2)$,
\begin{eqnarray}
&&M_{N,T}(t, 2) = \sum_{j_{1}, j_{2}, j_{3}, j_{4}}
\Big\{ 
\langle H_{j_{1} j_{2}} H_{j_{2} j_{3}} \rangle
\langle H_{j_{3} j_{4}} H_{j_{4} j_{1}} \rangle \nonumber\\
&&
+\langle H_{j_{1} j_{2}} H_{j_{3} j_{4}} \rangle
\langle H_{j_{2} j_{3}} H_{j_{4} j_{1}} \rangle 
+\langle H_{j_{1} j_{2}} H_{j_{4} j_{1}} \rangle
\langle H_{j_{2} j_{3}} H_{j_{3} j_{4}} \rangle
\Big\}, \nonumber
\end{eqnarray}
which has $(2\times 2-1)!!=3$ terms in the summand of 
$j_{1}, \cdots, j_{4}$. Substitution of (\ref{eqn:Wick2})
and binomial expansion give the $3 \times 2^{2}=12$
terms in the form,
$$
  M_{N,T}(t, 2)= \left( \frac{c^2}{2} \right)^2
\sum_{\ell=1}^{12} L_{\ell}
$$
with
\begin{widetext}
\begin{eqnarray}
  L_{1}&=& \sum_{j_{1}, j_{2}, j_{3}, j_{4}}
\delta_{j_1 j_3} \delta_{j_2 j_2} \delta_{j_3 j_1} 
\delta_{j_4 j_4}, \quad
  L_{2}= \sum_{j_{1}, j_{2}, j_{3}, j_{4}}
\delta_{j_1 j_3} \delta_{j_2 j_2} \delta_{j_3 j_4} 
\delta_{j_4 j_1} \gamma,
\nonumber\\
  L_{3}&=& \sum_{j_{1}, j_{2}, j_{3}, j_{4}}
\delta_{j_1 j_2} \delta_{j_2 j_3} \delta_{j_3 j_1} 
\delta_{j_4 j_4} \gamma, \quad
  L_{4}= \sum_{j_{1}, j_{2}, j_{3}, j_{4}}
\delta_{j_1 j_2} \delta_{j_2 j_3} \delta_{j_3 j_4} 
\delta_{j_4 j_1} \gamma^2,
\nonumber\\
  L_{5}&=& \sum_{j_{1}, j_{2}, j_{3}, j_{4}}
\delta_{j_1 j_4} \delta_{j_2 j_3} \delta_{j_2 j_1} 
\delta_{j_3 j_4}, \quad
  L_{6}= \sum_{j_{1}, j_{2}, j_{3}, j_{4}}
\delta_{j_1 j_4} \delta_{j_2 j_3} \delta_{j_2 j_4} 
\delta_{j_3 j_1} \gamma,
\nonumber\\
  L_{7}&=& \sum_{j_{1}, j_{2}, j_{3}, j_{4}}
\delta_{j_1 j_3} \delta_{j_2 j_4} \delta_{j_2 j_1} 
\delta_{j_3 j_4} \gamma, \quad
  L_{8}= \sum_{j_{1}, j_{2}, j_{3}, j_{4}}
\delta_{j_1 j_3} \delta_{j_2 j_4} \delta_{j_2 j_4} 
\delta_{j_3 j_1} \gamma^2,
\nonumber\\
  L_{9}&=& \sum_{j_{1}, j_{2}, j_{3}, j_{4}}
\delta_{j_1 j_1} \delta_{j_2 j_4} \delta_{j_2 j_4} 
\delta_{j_3 j_3}, \quad
  L_{10}= \sum_{j_{1}, j_{2}, j_{3}, j_{4}}
\delta_{j_1 j_1} \delta_{j_2 j_4} \delta_{j_2 j_3} 
\delta_{j_3 j_4} \gamma,
\nonumber\\
  L_{11}&=& \sum_{j_{1}, j_{2}, j_{3}, j_{4}}
\delta_{j_1 j_4} \delta_{j_2 j_1} \delta_{j_2 j_4} 
\delta_{j_3 j_3} \gamma, \quad
  L_{12}= \sum_{j_{1}, j_{2}, j_{3}, j_{4}}
\delta_{j_1 j_4} \delta_{j_2 j_1} \delta_{j_2 j_3} 
\delta_{j_3 j_4} \gamma^2.
\nonumber
\end{eqnarray}
\begin{figure}
\includegraphics[width=.6\linewidth]{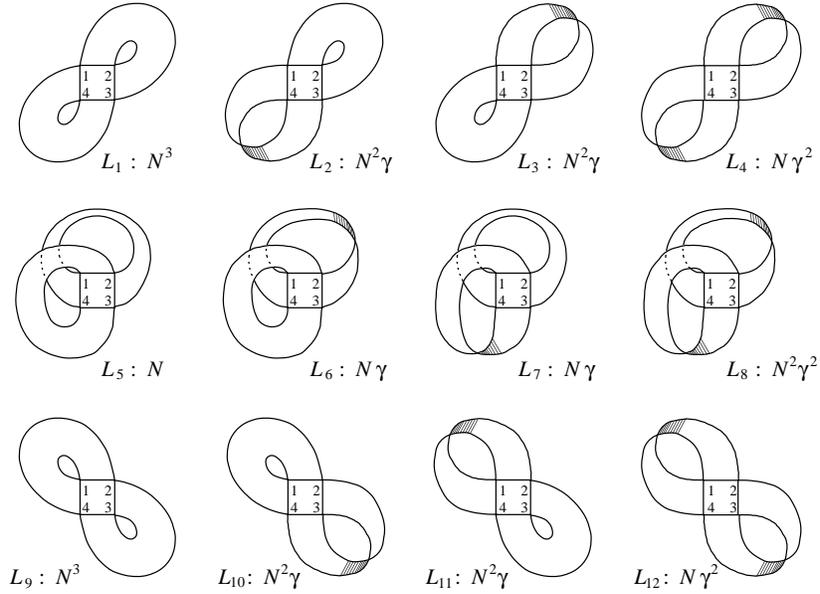}
\caption{M\"obius graphs for the fourth moment.\label{fig:fig1}}
\end{figure}
\end{widetext}

Graphically, we prepare a square with four vertices labeled 
$j_{1}, j_{2}, j_{3}, j_{4}$ in a cyclic order
for each term as shown in
Fig.1 and connect the vertices  $j_{a}$ and
$j_{b}$ by a line for each Kronecker's delta 
$\delta_{j_{a} j_{b}}$.
We then regard these lines connecting vertices
as the hems of ribbons connecting the two edges
of the square.
For example, the two lines connecting $j_1 \leftrightarrow j_3$
and $j_2 \leftrightarrow j_2$ in the term $L_{2}$ are considered
as the two hems of a ribbon, say $r_{1}$, connecting the edges
$\overline{j_1 j_2}$ and $\overline{j_2 j_3}$ of the square,
while the lines $j_3 \leftrightarrow j_4$ and
$j_4 \leftrightarrow j_1$ are as those of a ribbon, say $r_{2}$,
connecting $\overline{j_3 j_4}$ and $\overline{j_4 j_1}$.
There are two ways to connect two distinct edges by a ribbon,
by untwisting as $r_{1}$
and by twisting as $r_{2}$ in the above example,
respectively. For each twisted ribbon, we put a factor
$\gamma$.

Next we take the summation over $j_{1}, \cdots, j_{4}$ in each
term. Again consider the term $L_{2}$ for example.
Under the restrictions on indices specified by the
Kronecker deltas,
$j_{1}=j_{3}=j_{4}$, only two indices, say $j_{1}$ and $j_{2}$,
can be chosen arbitrary from $\{1,2, \cdots, N\}$.
Then the summation over all possible choices
of indices gives $N^2$ for this term.
The contribution of $L_{2}$ is thus $N^2 \gamma$.
We list up the contributions of all terms in Fig.1, and the
sum of them gives 
\begin{eqnarray}
&& M_{N,T}(t, 2)/ (c^2/2)^2 \nonumber\\
&=& (N^3+N^2 \gamma \times 2+ N \gamma^2) \times 2
+N+N \gamma \times 2 + N^2 \gamma^2 \nonumber\\
&=&N^3 \left\{ 
2+(4\gamma+\gamma^2) \frac{1}{N} 
+(1+2 \gamma + 2 \gamma^2) \frac{1}{N^2} \right\}.
\label{eqn:ex1}
\end{eqnarray}
This shows that the 12 terms are classified into 6 equivalence
classes, $\{L_{1}, L_{9}\}$,  $\{L_{2}, L_{3}, L_{10}, L_{11}\}$,
$\{L_{4}, L_{12}\}$, $\{L_{5}\}$, $\{L_{6}, L_{7}\}$ and
$\{L_{8}\}$, with respect to the contribution to
the moment, and Fig.1 implies that all graphs for the
terms in an equivalence class are topologically equivalent.

\subsection{General formula}

For the general $2k$-th moment, $M_{N,T}(t, k)$, $k \geq 1$,
we have $(2k-1)!!$ Wick couplings in the formula (\ref{eqn:Mk4}),
each term of which is the $k$ products of the variances
$\langle H_{j \ell} H_{m n}\rangle$. 
By applying (\ref{eqn:Wick2}) and expanding in $\gamma$,
we will have the $K=(2k-1)!! \times 2^k$ terms,
$M_{N,T}(t, k)/(c^2/2)^k=\sum_{\ell=1}^{K} L_{\ell}$.
As demonstrated in the above section, one-to-one
correspondence is established between terms $\{ L_{\ell} \}$
and graphs each of which consists of a $2k$-gon with its edges
connected by $k$ ribbons to each other. For each graph 
corresponding to $L_{\ell}$, let $\varphi_{\ell}$ be the
number of twisted ribbons in the $k$ ribbons
and $V_{\ell}$ be the ``free indices" remaining
after the identification of indices under the Kronecker delta
conditions. Then the contribution from the term $L_{\ell}$
is given by $N^{V_{\ell}} \gamma^{\varphi_{\ell}}$.
As mentioned at the end of the previous section,
we consider the equivalence classes of the terms
having the same contribution to the $2k$-th moment,
and let each equivalence class be represented by a graph $\Gamma$.
We let $V(\Gamma)$ and $\varphi(\Gamma)$ be the numbers of free indices
and of twisted ribbons. Moreover, we denote the number of elements
in the equivalence class $\Gamma$ by $|\Gamma|$.
In other words, $|\Gamma|$ is the number of ways
to generate graphs, which are topologically equivalent with 
$\Gamma$, using a $2k$-gon and $k$ ribbons
by gluing edges of $2k$-gons by ribbons.
Define ${\cal G}(k)$ be the
collection of all graphs $\{\Gamma\}$ generated
by the present procedure. Then we have
\begin{equation}
M_{N,T}(t, k)=
\left(\frac{c^2}{2}\right)^{k} 
\sum_{\Gamma \in {\cal G}(k)} 
|\Gamma| N^{V(\Gamma)} \gamma^{\varphi(\Gamma)}.
\label{eqn:Mkc}
\end{equation}

Each graph $\Gamma$ having no twisted ribbons, $\varphi(\Gamma)=0$,
defines a way of drawing a graph on an 
orientable surface $S_{\Gamma}$,
called a {\it map}, and each map specifies the surface
$S_{\Gamma}$ on which the graph is drawn
(see, for example, \cite{Zvo97,Oko99}).
In general, the specified orientable surface has
``holes" or ``handles" and their number is called the
{\it genus} $g(S_{\Gamma})$.
The genus is related with $V(\Gamma)$,
$E(\Gamma)=k$ (the number of distinct edges;
the original $2k$ sides of polygon were glued together in
pairs by ribbons) and 
$F(\Gamma)=1$ (the number of faces) through the Euler characteristic,
\begin{equation}
\chi(S_{\Gamma}) \equiv V(\Gamma)-k+1
=2-2g(S_{\Gamma}).
\label{eqn:Euler1}
\end{equation}
Then the contribution from all graphs having no twisted ribbons
is expressed as
\begin{eqnarray}
M_{N,T}^{0}(t, k) &=& \left(\frac{c^2}{2}\right)^k
\sum_{\Gamma \in {\cal G}(k)} {\bf 1}_{\{\varphi(\Gamma)=0\}}
|\Gamma|
N^{k+1-2g(S_{\Gamma})} \nonumber\\
&=& \left(\frac{c^2}{2}\right)^k N^{k+1}
\sum_{g=0}^{[k/2]} \varepsilon_{g}(k)
\left(\frac{1}{N^2}\right)^{g},
\label{eqn:Mkc0}
\end{eqnarray}
where 
${\bf 1}_{\{\omega\}}$ is the indicator; 
${\bf 1}_{\{\omega\}}=1$
if the condition $\omega$ is satisfied 
and ${\bf 1}_{\{\omega\}}=0$ otherwise, and
$\varepsilon_{g}(k)= \sum_{\Gamma \in {\cal G}(k)}
{\bf 1}_{\{\varphi(\Gamma)=0,
V(\Gamma)=k+1-2g\}} |\Gamma|$.

In the similar way, the graphs $\Gamma$ having twisted ribbons,
$\varphi(\Gamma) \geq 1$, are considered to define
non-orientable surfaces $S_{\Gamma}$.
The genus $g$ for non-orientable surface may be 
defined by the Euler characteristics as \cite{Thu97}
$$
\chi(S_{\Gamma})=2-g(S_{\Gamma})
$$
instead of (\ref{eqn:Euler1}). Then all the contribution
to the moment from such non-orientable surface graphs is
\begin{eqnarray}
&& M_{N,T}^{1}(t, k) \nonumber\\
&& = \left(\frac{c^2}{2}\right)^k
\sum_{\Gamma \in {\cal G}(k)} 
{\bf 1}_{\{\varphi(\Gamma) \geq 1\}} |\Gamma|
N^{k+1-g(S_{\Gamma})} \gamma^{\varphi(\Gamma)} \nonumber\\
&& = \left(\frac{c^2}{2}\right)^k N^{k+1}
\sum_{g=1}^{k} \left(\frac{1}{N}\right)^{g}
\sum_{m=1}^k \widetilde{\varepsilon}_{g,m}(k)
\gamma^{m},
\label{eqn:Mkc1}
\end{eqnarray}
where
$\widetilde{\varepsilon}_{g,m}(k)
=\sum_{\Gamma \in {\cal G}(k)} 
{\bf 1}_{\{\varphi(\Gamma)=m, V(\Gamma)=k+1-g\}} |\Gamma|$.
The moment (\ref{eqn:Mkc}) is then given by the summation
\begin{equation}
M_{N,T}(t,k)=M_{N,T}^{0}(t,k)+M_{N,T}^{1}(t,k).
\label{eqn:Mkcs}
\end{equation}

It should be noted that $\gamma$ defined by (\ref{eqn:gamma})
is a monotonically increasing function of $t$ and 
changes its value from $\gamma=0$ to 1 
as the time passes from $t=0$ to $T$. 
The above formula (\ref{eqn:Mkc1})
shows the fact that the contribution from M\"obius graphs
with twisted ribbons is growing in time and at $t=T$
twisted ribbons contribute with the same weights as
untwisted ribbons (the GOE case).

\section{CALCULATION BY DENSITY FUNCTION}

Set
\begin{eqnarray}
&& \rho_{N,T}(t, \x; T, \y)= \frac{C}{(N!)^2}
e^{-|\x|^2/2t} h_{N}(\x) {\rm sgn}(h_{N}(\y)) \nonumber\\
&& \quad \times 
\det_{1 \leq j, k \leq N} \left(
\frac{e^{-(x_{j}-y_{k})^2/2(T-t)}}{\sqrt{2 \pi (T-t)}} \right)
\nonumber
\end{eqnarray}
for $\x, \y \in \R^{N}$.
It is easy to confirm that
$\rho_{N,T}(t,\x; T, \y)$ is invariant under any
permutation of $x_{1}, \cdots, x_{N}$ and that
of $y_{1}, \cdots, y_{N}$, and (\ref{eqn:P0NT1}) is equal to
$N ! \int d \y \ \rho_{N,T}(t,\x; T, \y)$,
if $\x \in \R^{N}_{<}$.
Then the density function at time $t$ is defined as
\begin{equation}
\rho(t,x)= N \int \prod_{j=2}^{N} dx_{j} \int d \y \
\rho_{N,T}(t, \x; T, \y),
\label{eqn:R10}
\end{equation}
and the $2k$-th moment is given by
\begin{equation}
M_{N,T}(t, k) = \int x^{2k} \rho(t,x) dx.
\label{eqn:MkB1}
\end{equation}
Let $H_{j}(x)$ be the $j$-th Hermitian polynomial,
satisfying the orthogonality
$
  \int e^{-x^2} H_{j}(x) H_{\ell}(x) dx =h_{j} \delta_{j \ell} 
$
with $h_{j}=2^{j} j ! \sqrt{\pi}$.
As shown in Appendix A, the general formula for the
dynamical correlation functions of vicious walks
reported in \cite{NKT03,KNT03} gives the expression
\begin{eqnarray}
&& \rho(t,x) = \frac{1}{c} e^{-(x/c)^2} \sum_{j=0}^{N-1}
\frac{1}{h_{j}} \left\{ H_{j}(x/c) \right\}^{2}
\nonumber\\
&& + \frac{\gamma}{2^{N-1} c \sqrt{\pi}} e^{-(x/c)^2}
H_{N-1}(x/c)  \nonumber\\
&& \times \frac{1}{(N/2-1)!}
\sum_{j=0}^{\infty} \frac{(N/2+j)!}{(N+2j+1)!}
\gamma^{j} H_{N+2j+1}(x/c). \qquad
\label{eqn:R10b}
\end{eqnarray}
Substituting (\ref{eqn:R10b}) into (\ref{eqn:MkB1}) and
replacing the integral variable $x$ by $y=x/c$ give
\begin{eqnarray}
&& M_{N,T}(t, k)/c^{2k} =
\sum_{j=1}^{N-1} \frac{1}{h_{j}} \int y^{2k} \left\{ 
H_{j}(y) \right\}^2 e^{-y^2} dy \nonumber\\
&& + \frac{\gamma}{2^{N-1} \sqrt{\pi} (N/2-1)!} 
\sum_{j=0}^{\infty} \frac{(N/2+j)!}{(N+2j+1)!}
\gamma^{j} \nonumber\\
&& \times \int y^{2k} H_{N-1}(y) H_{N+2j+1}(y) 
e^{-y^2}dy \nonumber\\
&=& \frac{1}{2^{2k+N} \sqrt{\pi}(N-1)!} 
\sum_{j=0}^{k} \frac{(2k)!}{j! (2k-2j)!} \nonumber\\
&& \times \left[
\int H_{2k-2j}(y) \{H_{N}(y)\}^2 e^{-y^2} dy \right.
\nonumber\\
&& \quad \left.
- \int H_{2k-2j}(y) H_{N-1}(y) H_{N+1}(y) e^{-y^2} dy \right]
\nonumber\\
&& + \frac{\gamma}{2^{2k+N-1}\sqrt{\pi} (N/2-1)!}
\sum_{j=0}^{\infty} \frac{(N/2+j)!}{(N+2j+1)!}
\gamma^{j} \nonumber\\
&& \times \sum_{\ell=0}^{k} \frac{(2k)!}{\ell! (2k-2\ell)!} 
\nonumber\\
&& \times 
\int H_{2k-2\ell}(y) H_{N-1}(y) H_{N+2j+1}(y) e^{-y^2} dy.
\nonumber
\end{eqnarray}
In the second equality, we used the Christoffel-Darboux 
formula (see page 193 in \cite{Bat53})
\begin{eqnarray}
&&\sum_{j=0}^{N-1} \frac{1}{h_{j}} \{H_{j}(y)\}^2
\nonumber\\
&& =
\frac{1}{2^{N} \sqrt{\pi} (N-1)!} \left[
\{H_{N}(y)\}^2-H_{N-1}(y) H_{N+1}(y) \right],
\nonumber
\label{eqn:CD}
\end{eqnarray}
and the relation
$$
(2y)^{2k} = \sum_{j=0}^{k} 
\frac{(2k)!}{j! (2k-2j)!} H_{2k-2j}(y).
$$
Now we apply the integration formula for the triple
of Hermitian polynomials (see page 290 in \cite{Bat54})
\begin{widetext}
$$
\int H_{j}(y) H_{\ell}(y) H_{m}(y) e^{-y^2} dy
=
\left\{
   \begin{array}{ll}
      \displaystyle{
      \frac{j ! \ell ! m ! }
      {(s-j)! (s-\ell)! (s-m)!} 2^{s} \sqrt{\pi},
      } & 
   \mbox{if} \ j+\ell+m=2s \ \mbox{is even} \\
        &
        \mbox{and} \ s-j \geq 0, s-\ell \geq 0, s-m \geq 0, \\
        & \nonumber\\
      0, & \mbox{otherwise.} \nonumber\\
   \end{array}\right. 
$$
Then we arrive at
$M_{N, T}(t, k) = 
 M_{N,T}^{0}(t, k)+ M_{N,T}^{1}(t, k)$, with
\begin{eqnarray}
\label{eqn:Mkd0}
M_{N,T}^{0}(t, k) &=& \left(\frac{c^2}{2}\right)^k
\frac{(2k)!}{2^{k} k!} \sum_{j=0}^{k}
{k \choose j} {N \choose k-j+1} 2^{k-j} \\
\nonumber\\ 
\label{eqn:Mkd1}
M_{N,T}^{1}(t, k) &=& \left(\frac{c^2}{2}\right)^k
\sum_{j=0}^{k-1} \sum_{\ell=0}^{k-j-1}
\frac{2^{\ell-j}(N/2+\ell)! N! (2k)!}
{(N/2)! (N+\ell-k+j)!
(k-j-\ell-1)!(k-j+\ell+1)! j!} \gamma^{\ell+1}.
\end{eqnarray}
\end{widetext}
Here we do not repeat the explanation how to characterize
the quantity $\varepsilon_{g}(k)$ in
(\ref{eqn:Mkc0}) by using the formula (\ref{eqn:Mkd0}), and
only mention that it is obtained as the solution of the
recurrence relation
\begin{eqnarray}
&& (k+1)\varepsilon_{g}(k) =
(4k-2) \varepsilon_{g}(k-1) \nonumber\\
&& \quad +(k-1)(2k-1)(2k-3) \varepsilon_{g-1}(k-2)
\label{eqn:rec}
\end{eqnarray}
with the boundary conditions
$$
\varepsilon_{g}(0)
=
\left\{
   \begin{array}{ll}
      1, & 
   \mbox{if} \quad g=0, \\
      0, & \mbox{otherwise.} \nonumber\\
   \end{array}\right. 
$$
See \cite{HZ86}, Sec.5.5 in \cite{Meh91}
and \cite{Zvo97} for details.

In order to express the expansion in $1/N$ for (\ref{eqn:Mkd1}), 
here we introduce the number $s(n,k)$ defined as the 
coefficients of the expansion
$$
x(x-1) \cdots (x-n+1)=\sum_{\ell=1}^{n}s(n, \ell) x^{\ell}
$$
for $n \geq 1$. It is known that $s(n,\ell) \times (-1)^{n-\ell}$
is the number of elements in the set $S_{n}$ of all permutations
of $\{1,2, \cdots, n\}$, which are products of $\ell$ disjoint cycles.
These numbers $s(n, \ell)$ 
are called {\it the Stirling numbers of the
first kind} \cite{AS65}. 
For example, there are $3!=6$ distinct permutations of
$\{1,2,3\}$. We denote a permutation $1 \to a, 2 \to b, 
3 \to c$ simply by $[a \, b \, c]$
$(a,b,c \in \{1,2,3\}, a \not= b \not= c)$.
We regard $[2 \, 3 \, 1]$ as a cycle 
$1 \to 2 \to 3 \to 1$, $[2 \, 1 \, 3]$ as 
a product of two cycles $1 \to 2 \to 1$ and $3 \to 3$,
and the identity transformation $[1 \, 2 \, 3]$ as
that of three cycles $1 \to 1, 2 \to 2$ and $3 \to 3$.
In this example, we see $s(3,1)=2$ 
(for there are two elements $[2 \, 3 \, 1]$ and
$[3 \, 1 \, 2]$ with $\ell=1$ in $S_{3}$), 
$s(3,3)=1$ ($[1 \, 2 \, 3]$) and
$s(3,2)=-3$ (other three permutations).
For convenience, we will assume here that $s(n, \ell)=0$ if 
$n \leq 0$, or $\ell \leq 0$
or $n < \ell$. 
Then we have expression (\ref{eqn:Mkc1}) from (\ref{eqn:Mkd1}) 
with the coefficients
\begin{eqnarray}
&& \widetilde{\varepsilon}_{g, m}(k) =
\frac{(2k)!}{2^{k}} \sum_{j=k-g+1}^{k}
\sum_{\ell} \frac{ (-1)^{m-\ell} 2^{m+j-\ell}}
{(k-j)! (j-m)! (j+m)!} \nonumber\\
&& \quad \times s(m,\ell) s(j-m+1, k-g-\ell+2). 
\label{eqn:E}
\end{eqnarray}

It is easy to confirm that $s(n,1)=(-1)^{n-1} (n-1)!$,
$s(n,n-1)=-n(n-1)/2$ and $s(n,n)=1$ for any $n=1,2,3, \cdots$,
and numerical tables of $s(n, \ell)$ are found in \cite{AS65}.
For example, if we set $k=2$, (\ref{eqn:E}) gives
$\widetilde{\varepsilon}_{1,1}(2)=4 s(1,1)s(2,2)=4$,
$\widetilde{\varepsilon}_{1,2}(2)=s(2,2)s(1,1)=1$,
$\widetilde{\varepsilon}_{2,1}(2)=6 \left\{
s(1,1)s(1,1)+2s(1,1)s(2,1)/3 \right\}=2$
and
$\widetilde{\varepsilon}_{2,2}(2)=
-2s(2,1)s(1,1)=2$, and having
$\varepsilon_{0}(2)=2, \varepsilon_{1}(2)=1$,
the result (\ref{eqn:ex1}) is again obtained
through the general formula (\ref{eqn:Mkcs}) with
(\ref{eqn:Mkc0}) and (\ref{eqn:Mkc1}).

\section{CONCLUDING REMARKS}

In the present paper we performed the graphical expansions
for the moments of positions of vicious walkers.
The obtained formula (\ref{eqn:Mkcs}) with
(\ref{eqn:Mkc0}) and (\ref{eqn:Mkc1}) can be regarded as
the power series of the number of walkers $N$.
Here we consider the large $N$ limit. Let
$\hat{M}_{N,T}(t,k)$ be the dominant term of 
(\ref{eqn:Mkcs}) in $N \gg 1$. Then
\begin{equation}
\hat{M}_{N,T}(t,k) =\left(\frac{c^2}{2}\right)^{k} N^{k+1}
\varepsilon_{0}(k),
\label{eqn:Mhat}
\end{equation}
that is, only the contribution from the genus-zero
orientable surfaces will survives in the limit $N \to \infty$.
It is well known that $\varepsilon_{0}(k)$'s are given by
the Catalan numbers $C_{k}$ (see, for example, \cite{Slo73})
\begin{equation}
\varepsilon_{0}(k)=C_{k} = \frac{1}{k+1}
{2k \choose k},
\label{eqn:Cat1}
\end{equation}
and their generating function is
\begin{eqnarray}
a(\zeta) &=& \frac{1}{2\zeta} \left\{1-2\zeta-\sqrt{1-4\zeta}\right\}
\nonumber\\
&=& \sum_{k=1}^{\infty} \zeta^{k} C_{k}.
\label{eqn:Cat2}
\end{eqnarray}
Corresponding to (\ref{eqn:Mhat}), define the density function
$\hat{\rho}(t,x)$ as
$$
\hat{M}_{N,T}(t,k) = \int x^{2k} \hat{\rho}(t,x) dx.
$$
Multiplying the both sides by $z^{k}$ and taking
the summation over $k$ from 0 to $\infty$, we will have
$$
\frac{1}{c^2 z} \left[1-\sqrt{1-2c^2 N z} \right]
= \int \frac{\hat{\rho}(t,x)}{1-z x^2} dx,
$$
where (\ref{eqn:Mhat}), (\ref{eqn:Cat1}) and (\ref{eqn:Cat2})
were used. If we set $z=1/2c^2 N$, it becomes
$$
\int \frac{\hat{\rho}(t,x)}{2N-(x/c)^2} dx
=1.
$$
This integral equation can be solved as \cite{Meh91}
$$
\hat{\rho}(t,x)=
\left\{
   \begin{array}{ll}
      \displaystyle{
      \frac{1}{\pi c} \sqrt{2N-(x/c)^2},
      } & 
   \mbox{if} \ |x| \leq \sqrt{2N} c \\ 
       &  \\
      0, & \mbox{otherwise.} \nonumber\\
   \end{array}\right. 
$$
In the large $N$ limit, the density function will keep
a semicircle shape independently of the time evolution,
in which only the width of the semicircle depends on time
and simply scaled by $c$ \cite{NKT03}.
In other words, {\it Wigner's semicircle law} is universal
in $N \to \infty$.

The universal property of random matrix theory at the large $N$
limit and its finite-$N$ corrections have been studied by
calculating the Green functions in the form of $1/N$ expansions
in the field theory \cite{tHo74,BIPZ78,Cic82,Itoi97}.
In order to compare our present results with the previous
ones, here we consider the one-point Green function
defined by
\begin{equation}
G_{N, T}(t,z)=\left\langle \frac{1}{N} \sum_{j=1}^{N}
\frac{1}{z-\xi_{j}} \right\rangle_{t}
\label{eqn:Green0}
\end{equation}
with $\xi_{j}=x_{j}/(\sqrt{N} c)$ for $z \in {\bf C}$.
It is nothing but the generating function of the moments
(\ref{eqn:moment0}),
$$
G_{N, T}(t,z)=\frac{1}{Nz} \sum_{k=0}^{\infty}\
\left(\frac{1}{Nc^2 z^2}\right)^{k}
M_{N, T}(t,k).
$$
We define the coefficients $\bar{M}_{T}^{(n)}(t,k)$
in the expansion of the moment as
\begin{equation}
M_{N,T}(t,k)=\left(\frac{c^2}{2}\right)^{k}
N^{k+1} \sum_{n=0}^{\infty} \frac{1}{N^n}
\bar{M}_{T}^{(n)}(t,k).
\label{eqn:Mbar}
\end{equation}
Then we have the $1/N$ expansion of the Green function
as
\begin{equation}
G_{N, T}(t,z)=\sum_{n=0}^{\infty}
\frac{1}{N^n} G_{T}^{(n)}(t,z),
\label{eqn:Green1}
\end{equation}
where
\begin{equation}
G_{T}^{(n)}(t,z)=\frac{1}{z} \sum_{k=0}^{\infty}
\left(\frac{1}{2z^2} \right)^{k}
\bar{M}_{T}^{(n)}(t,k).
\label{eqn:GTn}
\end{equation}
Since we have obtained the closed expressions for any
moments, (\ref{eqn:Mkcs}) with (\ref{eqn:Mkd0}) and
(\ref{eqn:Mkd1}), the Green function (\ref{eqn:Green0})
is completely determined, and the coefficients 
$G_{T}^{(n)}(t,z)$ in its $1/N$-expansion (\ref{eqn:Green1})
can be derived for any order $n$ through the
formulae (\ref{eqn:Mkc0}), (\ref{eqn:Mkc1}) with
the known result of $\varepsilon_{g}(k)$
\cite{HZ86} and (\ref{eqn:E}).
For example, as shown in Appendix B, the present results give 
the following expressions for the first three terms;
\begin{eqnarray}
\label{eqn:G0}
G_{T}^{(0)}(t,z) &=& \frac{1+a(\zeta)}{z}, \\
\label{eqn:G1}
G_{T}^{(1)}(t,z) &=& 2z \frac{a(\zeta)}{1-a(\zeta)^2} \times
\frac{\gamma a(\zeta)}{1-\gamma a(\zeta)}, 
\end{eqnarray}
and
\begin{eqnarray}
&& G_{T}^{(2)}(t,z) = \frac{\zeta^2}{3 z} 
\frac{\partial^2}{\partial \zeta^2} 
\frac{1}{1-a(\zeta)^2} + \frac{\zeta}{6 z} 
\frac{\partial}{\partial \zeta} 
\frac{1}{1-a(\zeta)^2}
\nonumber\\
&& \qquad +\frac{1}{2z} \frac{\partial}{\partial \zeta} \left[
\frac{a(\zeta)}{1-a(\zeta)^2} 
\left\{ 2 \gamma \frac{\partial}{\partial \gamma}-1 \right\}
\frac{\gamma a(\zeta)}{1-\gamma a(\zeta)} \right] \nonumber\\
&& \qquad - \frac{1}{2 z \zeta} \frac{a(\zeta)}{1-a(\zeta)^2}
\left\{ 3 \gamma \frac{\partial}{\partial \gamma}-1 \right\}
\frac{\gamma a(\zeta)}{1-\gamma a(\zeta)}
\label{eqn:G2}
\end{eqnarray}
with $\zeta=1/(2z^2)$.
By using (\ref{eqn:Cat2}), we can show that
setting $\gamma=1$ reduces them to 
Equations (24), (26) and (28) with $\beta=1$ (the GOE case)
given by Itoi, respectively, 
who derived them by solving the loop equations \cite{Itoi97}.
On the other hand, we can confirm that, if we set $\gamma=0$,
$G_{T}^{(n)}(t,z/\sqrt{2})/\sqrt{2}$ gives the $\beta=2$
(GUE) results of Itoi. Some details are given in Appendix B.
Our result is new and general, and it will reproduce the
previous results of $1/N$-expansions for the GUE and
GOE as the special cases with $\gamma=0$ and 1,
respectively.

Our formula also gives a power series
in $\gamma$ with $N$-dependent coefficients.
Let $\widetilde{M}(k; \gamma, N)=M_{N,T}(t,k)/(c^2/2)^k$.
Then 
$$
  \widetilde{M}(k; \gamma, N)= \sum_{m=0}^{k} \gamma^{m}
f_{k, m}(N),
$$
where $f_{k, m}(N)$ are the polynomials of degree $k+1$ in $N$
for $m=0$ and of degree $k$ in $N$ for $1 \leq m \leq k$.
Explicit expression of $f_{k,m}(N)$ is immediately obtained
from (\ref{eqn:Mkd1}). We remark that the coefficients of the
highest order in $\gamma$, $f_{k,k}(N)$, is equal to
the {\it zonal polynomials} 
$Z_{(k)}(s_{1}, s_{2}, \cdots, s_{k})$, 
if we set all the variables
$s_{1}=s_{2}=\cdots=s_{k}=N$ \cite{MPH95}.

As shown by (\ref{eqn:P0NT3}) with (\ref{eqn:mu}),
the present system of vicious walkers can
be regarded as a Gaussian model
as matrix model, and thus exactly solvable as demonstrated
in this paper. We would like to state, however, that 
the probability density (\ref{eqn:P0NT1}) of the positions of 
walkers $\x$ is not in the simple Gaussian form multiplied
by $h_{N}(\x)$, when $0 < t < T$, due to the 
factor ${\cal N}_{N}(T-t, \x)$. Combination of the
Pfaffian representation of this factor given in
\cite{KT02a,KT02b} and the facts that
${\rm Pf}(A)=(\det A)^{1/2}$ for any even-dimensional
antisymmetric matrix $A$ and $\det A=e^{{\rm tr} \ln A}$,
(\ref{eqn:P0NT1}) is written for even $N$ as
$$
\rho_{N, T}(t, \x) \propto
h_{N}(\x) \exp \left(-V(\x) \right)
$$
with the non-harmonic potential
$$
V(\x)=-\frac{|\x|^2}{2t}+\frac{1}{2}
{\rm tr} \Big( \ln F(T-t, \x) \Big),
$$
where $F(s, \x)$ is the $N \times N$ antisymmetric matrix
with the element
$F_{jk}(s, \x)=(2/\sqrt{\pi})
{\rm Erf}((x_{k}-x_{j})/2\sqrt{s})$
with ${\rm Erf}(u)=\int_{0}^{x} du \, e^{-u^2}$.
Then the present results for $0 < t < T$ are nontrivial.

In summary, we demonstrated the GUE-to-GOE transition in time
for the vicious walk model by presenting the graphical 
expansion formula of the moments of walker's positions.
The weights of contributing graphs are time-developing
and our formula interpolates the genus expansion of
orientable graphs for GUE and that of non-orientable graphs
for GOE by using a time-parameter $\gamma=t/(2T-t)$.
The formula provides a power series 
in the number of walkers $N$ and
it was shown that the exact expression of dynamical correlation
functions recently reported by Nagao {\it et al.} \cite{NKT03}
is very useful in order to evaluate the coefficients
in the series. 
By comparing with the previous results
of $1/N$ expansion of the one-point Green function, 
we showed that our results are general and valid.
Further applications of the quaternion 
determinantal expressions of dynamical correlations
of vicious walkers to the graphical expansions
of more general types of moments
({\it e.g.} correlators of 
moments at different times \cite{IS97})
will be interesting future problems.

\appendix
\section{DENSITY FUNCTION}

Let $H_{j}(x)$ be the $j$-th Hermite polynomial defined
in Sec.IV.
We can read the density function (\ref{eqn:R10}) from
\cite{NKT03,KNT03} by setting $M=1$ as
\begin{equation}
\rho(t,x)=\sum_{\ell=0}^{N/2-1}
\frac{1}{r_{\ell}} \left[
\Phi_{2\ell}(x) R_{2\ell+1}(x)
-\Phi_{2\ell+1}(x) R_{2\ell}(x) \right],
\label{eqn:R10A}
\end{equation}
where
\begin{eqnarray}
\label{eqn:R}
R_{\ell}(x) &=& \gamma^{\ell/2} \sum_{j=0}^{\ell} \alpha_{\ell j}
H_{j}(x/c) \gamma^{-j/2}, \\
\Phi_{\ell}(x) &=& \int dy R_{\ell}(y) \nonumber\\
&& \times
\int_{-\infty < z < z' < \infty} dz dz'
\left| 
\matrix{ p(y,z) & p(x,z) \cr p(y, z') & p(x, z') } \right|, 
\nonumber\\
\label{eqn:r}
r_{\ell} &=& \frac{2}{\pi} \gamma^{2\ell+1/2} 
\left(\frac{c}{2}\right)^{4\ell+1} h_{2\ell},
\end{eqnarray}
with
\begin{eqnarray}
\alpha_{2\ell \, j} &=& (c/2)^{2\ell}
\delta_{2\ell \, j}, \nonumber\\
\alpha_{2\ell+1 \, j} &=& (c/2)^{2\ell+1}
(\delta_{2\ell+1 \, j} - 4\ell \delta_{2\ell-1 \, j} ),
\nonumber
\end{eqnarray}
and
$$
p(x,y)= \frac{e^{-x^2/2t}}{\sqrt{2 \pi t}} \times
\frac{e^{-(x-y)^2/2(T-t)}}{\sqrt{2 \pi (T-t)}}.
$$
The function $p(x,y)$ can be expanded using the
Hermite polynomials as
\begin{eqnarray}
 &&p(x,y) \nonumber\\
&=& \frac{e^{-(x/c)^2} e^{y^2/2T}}
{\sqrt{2 \pi t} \sqrt{2\pi(T-t)}}
\exp \left\{ -\frac{(y/\sqrt{T}-
\sqrt{\gamma}x/c)^2}{1-\gamma} \right\} \nonumber\\
&=& \frac{e^{-(x/c)^2} e^{-y^2/2T}}
{\sqrt{2 \pi t} \sqrt{2T-t}}
\sum_{j=0}^{\infty} \frac{\gamma^{j/2}}{h_{j}}
H_{j}(x/c) H_{j}(y/\sqrt{T}).
\nonumber
\end{eqnarray}
Then, for $\ell=0,1,2, \cdots$, we will have
\begin{eqnarray}
\Phi_{2\ell}(x) &=& \frac{r_{\ell}}{c^2} e^{-(x/c)^2} \nonumber\\
&\times&
  \sum_{j \geq 2\ell+1}
  \frac{\gamma^{\{j-(2\ell+1)\}/2}}{h_{j}}
  \beta_{j \, 2\ell+1} H_{j}(x/c), \nonumber\\
\Phi_{2\ell+1}(x) &=& 
-\frac{r_{\ell}}{c^2} e^{-(x/c)^2} \nonumber\\
&\times&
  \sum_{j \geq 2\ell}
  \frac{\gamma^{(j-2\ell)/2}}{h_{j}}
  \beta_{j \, 2\ell} H_{j}(x/c),
\label{eqn:Phi2}
\end{eqnarray}
where $\beta_{j\ell}$'s satisfy the relation
$$
\sum_{\ell=s}^{j} \beta_{j \ell} \alpha_{\ell s}=\delta_{js},
\quad 0 \leq s \leq j.
$$
Substituting (\ref{eqn:R}), (\ref{eqn:r}) and (\ref{eqn:Phi2})
into (\ref{eqn:R10A}) gives (\ref{eqn:R10b}).

It should be noted that, though the $M=1$ case of \cite{NKT03}
is equivalent with Pandey-Mehta's two-matrix model
\cite{PM83}, the present expression (\ref{eqn:R10b})
is more useful for the moment calculation 
as shown in Sec.IV.

\section{COEFFICIENTS IN $1/N$-EXPANSION OF 
GREEN FUNCTION}

By definition of $\bar{M}_{T}^{(n)}(t,k)$ in (\ref{eqn:Mbar}),
(\ref{eqn:Mkc0}) and (\ref{eqn:Mkc1}) give
\begin{eqnarray}
\label{eqn:Mbar0}
&& \bar{M}_{T}^{(0)}(t,k) =
\varepsilon_{0}(k)=C_{k}, \\
&& \bar{M}_{T}^{(1)}(t,k) =
\sum_{m=1}^{k} \widetilde{\varepsilon}_{1,m}(k)
\gamma^{m} \nonumber\\
\label{eqn:Mbar1}
&& \quad = \sum_{m \geq 1} {2k \choose k+m} \gamma^{m},\\
&& \bar{M}_{T}^{(2)}(t,k) = \varepsilon_{1}(k)
+ \sum_{m=1}^{k} \widetilde{\varepsilon}_{2,m}(k)
\gamma^{m} \nonumber\\
&& \quad = \frac{1}{12} \frac{(2k)!}{k! (k-2)!}
\nonumber\\
&& \quad + \frac{1}{2} \sum_{m \geq 1} \left\{
(2m-1) {2k \choose k+m} (k+1) \right. \nonumber\\
\label{eqn:Mbar2}
&& \qquad \qquad \qquad \left.
-(3m-1) {2k \choose k+m} \right\}, 
\end{eqnarray}
where we have used (\ref{eqn:E}), (\ref{eqn:Cat1}), and
the fact 
$\varepsilon_{1}(k)=(2k)!/\{12 k! (k-2)!\}$,
which is obtained from (\ref{eqn:rec}) with (\ref{eqn:Cat1}).
Through (\ref{eqn:Cat2}) and (\ref{eqn:GTn}), 
(\ref{eqn:Mbar0}) immediately gives
(\ref{eqn:G0}). In order to derive (\ref{eqn:G1}) and
(\ref{eqn:G2}) from (\ref{eqn:Mbar1}) and (\ref{eqn:Mbar2}),
respectively, we can use the identity given as Equation (C.5)
in \cite{KI97},
$$
\sum_{n=1}^{\infty} {2n \choose n+m} \zeta^{n+1}
= \frac{a(\zeta)^{m+1}}{1-a(\zeta)^2} 
\quad \mbox{for} \quad m \geq 0,
$$
and its derivatives with respect to $\zeta$.
It is easy to see that
$$
G_{T}^{(0)}(t,z)=z-\sqrt{z^2-2},
$$
and
$G_{T}^{(1)}(t,z)$ is zero at $\gamma=0$.
Moreover, we have obtained the following expressions for 
$0 \leq 1-\gamma \ll1$
\begin{widetext}
\begin{eqnarray}
&& G_{T}^{(1)}(t,z)=
-\frac{1}{2}\left[ \frac{1}{\sqrt{z^2-2}}
-\frac{z}{z^2-2} \right] 
-\frac{1}{2(z^2-2)^{3/2}} (1-\gamma)+{\cal O}((1-\gamma)^2),
\nonumber\\
&& G_{T}^{(2)}(t,z)=
\frac{2z^2+3-2z \sqrt{z^2-2}}
{4 (z^2-2)^{5/2}}
-\frac{z(z^2+6)-(z^2+1)\sqrt{z^2-2}}{4(z^2-2)^3}(1-\gamma)
+{\cal O}((1-\gamma)^2),
\nonumber
\end{eqnarray}
and 
$$
G_{T}^{(2)}(t,z)=\frac{1}{4 (z^2-2)^{5/2}}
\quad \mbox{at} \quad \gamma=0.
$$
\end{widetext}

\begin{acknowledgments}
The authors thank S. Kakei and T. Sasamoto for letting them know
the papers by M. Mulase. They also thank T. Nagao for useful comments
on the present work. They acknowledge the comments of a referee, which 
enabled them to show the calculations of one-point Green function
in the present paper.
\end{acknowledgments}

\end{document}